\documentclass[sn-mathphys]{sn-jnl}% Math and Physical Sciences Reference Style
\jyear{2021}%

\theoremstyle{thmstyleone}%
\theoremstyle{thmstyletwo}%

\theoremstyle{thmstylethree}%

\begin{document}

\title[Article Title]{High quality quasinormal modes of phononic crystals for quantum acoustodynamics}


\author*[1,2]{\fnm{Aleksey N.} \sur{Bolgar}}\email{alexgood@list.ru}

\author[2,3]{\fnm{Shtefan V.} \sur{Sanduleanu}}

\author[2,1]{\fnm{Aleksandr} \sur{Strelnikov}}

\author[1,2,4,5]{\fnm{Oleg V.} \sur{Astafiev}}

\affil[1]{\orgname{Skolkovo Institute of Science and Technology}, \orgaddress{\street{Bolshoy Boulevard 30, bld. 1}, \city{ Moscow}, \postcode{121205}, \country{ Russia}}}

\affil[2]{\orgdiv{Laboratory of Artificial Quantum Systems}, \orgname{Moscow Institute of Physics and Technology}, \orgaddress{\street{9 Institutskiy per.}, \city{ Dolgoprudny}, \postcode{141700}, \state{Moscow Region}, \country{Russia}}}

\affil[3]{\orgname{National University for Science and Technology (MISiS)}, \orgaddress{ \street{ Leninskiy Pereulok 4},\city{ Moscow}, \postcode{119049}, \country{Russia}}}

\affil[4]{\orgdiv{Physics Department}, \orgname{Royal Holloway, University of London}, \orgaddress{ \city{ Egham}, \postcode{ Surrey TW20 0EX}, \country{United Kingdom}}}

\affil[5]{\orgname{National Physical Laboratory}, \orgaddress{\street{Hampton Road}, \city{Teddington}, \postcode{Middlesex TW11 0LW}, \country{United Kingdom}}}


\abstract{Phononic crystals are a promising platform for the study of quantum acoustodynamics. In a recent experiment, the interaction of a superconducting quantum bit with modes of a phononic crystal has been demonstrated. The field of these modes is localized in a compact area, providing high values of the coupling constant with the qubit.
However, the Q-factor of  phononic crystal modes is strongly limited ($\sim$1050) due to a phonon emission from the crystal ends. For further use of phononic crystals in research in the field of quantum acoustodynamics, it is desirable to overcome this limitation in the quality factor.
In this work, we have proposed a structure consisting of a phononic crystal placed between the Bragg mirrors. Our simulations predict that the Q-factor in such a structure can reach ($\sim$100000). We demonstrate experimental results in which this structure has a Q-factor ($\sim$ 60,000), which is 60 times higher than that of an acoustic crystal of the same size.
}



\maketitle\
\section{Introduction}\label{sec1}\

  Quantum acoustodynamics  (QAD) studies the interaction of mechanical oscillation fields with quantum systems. This is a new direction of quantum mechanics, which was developed through the experimental study of hybrid systems in which artificial atoms interact with phonons \cite{gustafsson2014propagating, manenti2017circuit, bolgar2018quantum, o2010quantum, rouxinol2016measurements, lahaye2009nanomechanical, pirkkalainen2013hybrid}. The mechanical part of such systems can be represented by bulk resonators \cite{Cleland2010} or by surface acoustic wave (SAW) resonators \cite{manenti2016surface, gustafsson2014propagating}. The artificial atom in these systems is represented by superconducting qubits, which are the most promising for the development of quantum informatics \cite{arute2019quantum, preskill2018quantum, ye2019propagation, gu2017microwave, you2011atomic, kockum2019quantum} .

  SAW resonators for QAD are much easier to fabricate than bulk resonators, since they require only standard lithographic approaches that are commonly used to make devices with superconducting qubits. Since the length of acoustic waves is several orders less than the length of electromagnetic waves for the same frequencies, the device for QAD can be much more compact than a similar device for quantum electrodynamics (QED).
 
 A pioneering QAD device in Ref. \cite{gustafsson2014propagating} demonstrated the interaction of a transmon qubit with a traveling SAWs, which were not confined in a resonator. This achievement stimulated the experimental study of SAW - resonators. In Ref. \cite{manenti2016surface, magnusson2015surface} SAW resonators of gigahertz frequency range were shown (up to 4.4 GHz), which is important for their use in qubit readout schemes. Later, our group demonstrated the interaction of resonators of this type with an artificial atom in the quantum regime \cite{bolgar2018quantum}.

  In a typical SAW resonator, a standing mechanical wave is formed in a cavity between two Bragg mirrors. In Ref. \cite{bolgar2020phononic}, we show that the periodic structure of a long interdigital transducer (IDT) works like a phononic crystal. Quasinormal modes (QNMs) in such a structure are formed only due to the partial reflection of the SAW from the IDT strips. The resonant frequencies of quasinormal modes are slightly shifted in comparison to the resonant frequencies of the resonator with the same period and correspond to the acoustic branch of the phononic crystal. It does not have Bragg mirrors for the formation of standing mechanical waves. Nevertheless, the mechanical field of quasinormal modes in the phononic crystal is well localized, which makes it possible to achieve large values of a coupling strength with a quantum bit. In Ref. \cite{bolgar2020phononic} we demonstrated a strong coupling with an absolute coupling strength up to 39 MHz, which is 2.4 times greater than the previously demonstrated coupling with a typical SAW resonator \cite{bolgar2018quantum}. Thus, it was shown that acoustodynamic devices can have a simpler and more compact structure based on phononic crystals instead of SAW resonators. On the other hand, the quasinormal modes of a phononic crystal demonstrate much lower Q-factors than typical SAW resonators, since there is an energy leak through the acoustic wave radiation from the ends of the phononic crystal. In Ref. \cite{bolgar2020phononic}, the best Q-factor of the quasinormal mode is only 1050.  
  
  In this work, we propose a solution for improving the Q-factors and show that the Q-factor of its quasinormal modes can be significantly increased. We are studying phononic crystals with a nonuniform period. In this structure, two regions located at the ends of the crystal have a period of strips, that is slightly more, than in the central region of the crystal. Due to this special period these parts of the phononic crystal work as ideal Bragg mirrors for the fundamental quasinormal mode, concentrated in the central region of the device. That helps to reduce significantly the energy loss through SAW radiation from the crystal ends. Consequently, the Q-factors are increased. Numerical modeling of the acoustic field in our structures shows that Q-factors up to 100000 can be reached (for the modes near 3.1 GHz). They are limited by material propagation loss in that case \cite{manenti2016surface} . We also show the  results of experiments with a phononic crystal of the proposed type in the temperature range from room temperature to 15 mK. The maximum internal Q-factor at low temperatures reaches 61000.
  
     This article is organized as follows. First, we give an introduction into the energy dissipation channels in conventional SAW resonators and in phononic crystals in Sections 2 and 3, respectively. Our experimental device structure and measurement results are described in Section 4. Section 5 describes our method for modeling the acoustic field and theoretical results for phononic crystals of the considered type. Finally, our conclusions are presented in the final section 6.

\section{Energy losses description in a SAW resonator}\label{sec2}\

  It is important to compare structures of a typical SAW resonator and a phononic crystal in terms of their energy losses. The main mechanisms of the energy dissipation in these structures are different, which affects the Q-factors of these devices. We should describe these differences in order to improve later the characteristics of the phononic crystal while maintaining its advantages. 
   A typical SAW resonator is shown schematically in Fig.1a. It consists of two Bragg mirrors. Each mirror consists of an array of equidistant metallic strips of length $W$ , which are spaced by a uniform period $p'$. These stripes are formed on the surface of the substrate by metallization deposited through an organic mask prepared by electron beam lithography. The rate of SAW propagation is different on the free surface and under the metallized strips. This leads to partial SAW reflection from the boundary of the metallic strips with a reflection coefficient $r_s\propto\frac{h}{d}$, where $h$  is the thickness of metallization, and $\lambda$  is the SAW wavelength.
   
      Typically, this single strip reflectance is very low. For our devices on a quartz substrate $r_s\approx1.5\times 10^{-2}$. Nevertheless, for an array of strips with a period equal to the SAW wavelength, these partially reflected waves interfere constructively. That leads to an increase in the total value of the reflection coefficient of the Bragg Mirror  
$R=\tanh{(\mid r_s\mid N_g)}$, which is close to unity for mirrors consisting of a large number of strips $N_g$. These reflections form standing waves in a two-dimensional cavity between two Bragg mirrors. In fact, the set of normal modes of the SAW resonator is represented by the standing waves.

     There are 3 main channels of energy dissipation in such systems. The first one is the loss of energy through the mirrors, since their reflection coefficient ${R_m}$ cannot be equal to unity. Thus, the mirrors have a weak transparency $1 - {R_m}$, which provides the leakage of phonons accumulated in the SAW resonator. This radiation limits the Q-factor of the SAW resonator to a value ${Q_g}$ which for its fundamental mode with $\lambda _0 = p'$ is defined as \cite{morgan2010surface}

\begin{equation}
{Q_g} = \frac{\pi L_c}{\lambda _0 ( 1 - \tanh (\mid r_s \mid N_g)}, 
\label{eq1}
\end{equation}
where $L_c=d+\lambda_0/\mid 2r_s \mid$ is the effective length of the cavity, $d$ is the distance between the mirrors. Using (1) we get ${Q_g} \gg {10^7}$   for typical values \cite{manenti2016surface, Delsing2022PRX} $d \approx 100{\lambda _0},\,\,r \approx 0.02,\,\,{N_g} \approx 1000$. Nevertheless, experiments with such resonators on a quartz substrate demonstrate their internal Q-factors, which are two orders of magnitude lower \cite{manenti2016surface}. That means the main energy losses in such resonators are determined by other mechanisms. 

         The second dissipation channel is a diffraction loss. Due to the finite aperture of the mirrors $W$ , some of the reflected waves propagate at angles up to  ${\alpha _d} \approx \lambda /W$  and, therefore, leave the cavity of the resonator (see Fig. 1a). The diffraction contribution to the quality factor can be estimated as \cite{Aref2016}

\begin{equation}
Q_d = \frac{5\pi }{\mid 1 + \gamma \mid} \left( \frac{W}{\lambda _0} \right)^2,
\label{eq2}
\end{equation}
where $\gamma  = 0.378$ for ST-X quartz substrate at room temperature \cite{Aref2016}. Using (2) for the resonator widths $W = 100 - 150 \cdot \lambda _0$, we obtain ${Q_d}\sim 114 - 256 \cdot {10^3}$. In the experiment, resonators with such parameters at temperature 10  mK have demonstrated Q-factors \cite{manenti2016surface} ${Q_d}\sim 170 - 450 \cdot {10^3}.$   For a GaAs substrate at room temperature 300 K we obtain ${Q_d}\sim490 \cdot {10^3}$ by using (2) for $W = 120{\lambda _0}$. For such resonators, the Q-factors 
${Q_d}\sim 450 \cdot 10^3$  \cite{andersson2021acoustic} were demonstrated in a low-temperature experiment (10 mK). Thus, we can conclude that for resonators of this type, the Q-factor is mainly determined by diffraction losses.

\begin{figure}[h]%
\centering
\includegraphics[width=0.6\textwidth]{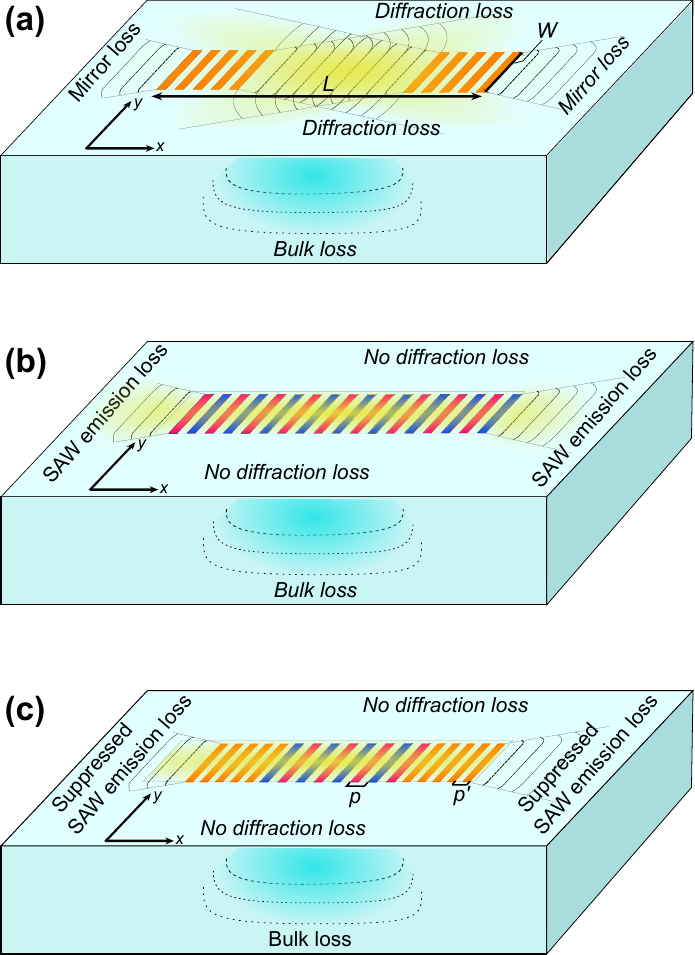}
\caption{\textbf{Energy   dissipation  in  a  SAW resonator  and  in  a  phononic  crystal.}
\textbf{a. SAW - resonator.} Two Bragg mirrors (orange strips) form a cavity for surface phonons on a piezoelectric substrate. Normal modes of such a resonator lose their energy due to diffraction losses, partial transparency of the mirrors, and losses in the substrate material. The approximate distribution of the SAW amplitude is shown with a yellow gradient. The bulk mode is shown with a blue gradient.
\textbf{b. Phononic crystal.} An array of  IDT metal strips (blue strips are grounded, red strips are connected to the microwave line) with a uniform period forms a medium with periodically modulating SAW propagation velocity. The intensity of the main quasinormal modes drops to zero (yellow gradient) in the Y-direction at the edges of the stripes. This eliminates diffraction losses. Thus, quasinormal modes lose their energy mainly due to the radiation of the SAW from the ends of the crystal.
\textbf{c. Modified phonon crystal.} The end parts (orange stripes) of a phononic crystal have a longer period and act as Bragg mirrors for the fundamental quasinormal mode of the crystal, which helps to conserve its energy.
}\label{fig1}
\end{figure}

The third channel of energy dissipation is the decay of a traveling SAW due to the excitation of mechanical modes in the bulk of the substrate. These material propagation losses provide an additional contribution to the quality factor $Q_m$, which can be estimated as \cite{manenti2016surface}

\begin{equation}
Q_m = \frac{\pi f_0l}{v},
\label{eq3}
\end{equation}

where $f_0$ is the SAW frequency, $v$  is the speed of SAW, $l$ is the phonon mean free path. From experiments on measuring a set of SAW resonators on ST-X quartz \cite{manenti2016surface}, it was found that ${Q_m} \sim 100 \cdot {10^3}$ for a frequency  of 3.1 Ghz  at 10 mK.

       Finally, the expression for the internal Q-factor of the SAW-resonator should use all the above contributions

\begin{equation}
Q_{i_{RES}} = ( \, \frac{1}{Q_g} + \frac{1}{Q_d}+ \frac{1}{Q_m}) \, ^{-1}.
\label{eq4}
\end{equation}

  In \cite{bolgar2018quantum}, our acoustodynamic device consisted of a superconducting qubit, which was coupled to a SAW resonator of the type described above. This device demonstrated the Q-factor of the resonator $Q_i\approx 9500$  and the coupling strength constant with the qubit $g\approx 16$ MHz. To develop more complex acoustodynamic devices, it is necessary to increase both of these characteristics simultaneously. In such devices with fixed optimal value of qubit transducer capacitance and varied values of  $L_c$  and $W$  the coupling constant is inversed proportional to the root of the cavity area   $g\propto (\,L_c W)\,^{-1/2}.$ Thus, we have to choose compromise values of the parameters $L_c$ and $W$ to suppress diffraction losses (2), and losses through mirrors (1), respectively. However, this will lead to suppression of the coupling constant value $g$. 
  
    We have managed to partially  resolve this contradiction  by replacing the typical  SAW resonator with a phononic crystal, as we did in Ref. \cite{bolgar2020phononic}.

\section{Energy losses description in a phononic crystal}\label{sec3}\

  The structure of a phononic crystal for SAW is shown schematically in Figure 1b. This is an array of metal strips, which are equidistantly deposited on a substrate with a period $p$. Due to the difference in the speed of SAW propagation on the free surface and under metallization, such a system is a medium with a periodically modulated refractive index. Therefore, SAWs in such a structure are reflected partially at the edge of each strip. These reflected waves interfere with each other, forming a set of unique eigenmodes for a given crystal. The distribution of the acoustic field of these modes depends on geometric parameters such as the period size $p$ , the metal strips length $W$ , their number $N$, and the ratio of the strip width to the period $\frac{a}{p}$,  which is usually close to 0.5. We describe a method for numerical simulation of the acoustic field distribution of quasinormal modes in the 5th section.
  
    Phononic crystals in the general case can have two main energy dissipation channels. First, these systems do not have Bragg mirrors, unlike the SAW resonator. Consequently, the contribution to the quality factor  ${Q_g}$ disappears, since it is determined by the partial transparency of the Bragg mirrors. Moreover, the acoustic field is not locked here, and SAWs can freely radiate from the ends of the crystal. The magnitude of that radiation leakage is determined by the geometrical parameters of the entire phononic crystal. This type of leakage is specific to a phononic crystal and it makes the main contribution to the Q-factor suppression for quasinormal modes. Therefore, we define the corresponding contribution to the quality factor as $Q_r$. Its value can be calculated from the modeling of the phononic crystal modes described in the 5th section. 
    
    Second, as in the case of a SAW resonator, the modes of a phononic crystal can excite bulk vibrations in the substrate, which leads to a contribution to the quality factor $Q_m$ due to material propagation losses. 
    
    Our approach is to localize the acoustic wave field in the crystal and minimize leakage along (SAW radiation from the ends) $1/Q_r$ and across (diffraction loss) the phononic crystal  $1/Q_d$. 
    
    An important advantage of a phononic crystal versus a SAW resonator is that its structure excludes diffraction losses in the fundamental modes due to their total reflection at the x-oriented crystal boundary. In our experimental samples, the effective filling factor of a phonon crystal by metallization of strips is about 60$\%$. That provides an effective refractive index in the region of the crystal $n_{eff}\approx 1.009$.  Thus, the total reflection angle at the grating boundary is  $\alpha_c=\arcsin(\,\frac{1}{n_{eff}})\,=82^\circ$.  We denote each quasinormal mode by two indices $i$ and $j$, which denote the orders of the modes along the x and y axes, respectively. Thus, the permitted values of the y-component of the wave vector are determined as  $k_y=\pi j/W$, where $j$ are integers. The angle of incidence $\alpha$ on the x-oriented boundary is defined as  $\alpha=\arctan(\,\frac{k_x}{k_y})\,$. Waves with $\alpha>\alpha_c$ are fully reflected from the x-oriented border. Taking $k_x\approx\frac{\pi}{p}$ for the main modes, and  $W = 12\,\mu$m we get 
$j\leq3$. These modes are free from diffraction losses, and the expression for their internal Q-factor should include only two main contributions

\begin{equation}
Q_{i_{PC}} = ( \, \frac{1}{Q_r} + \frac{1}{Q_m}) \, ^{-1}.
\label{eq5}
\end{equation}

   In our work in Ref. \cite{bolgar2020phononic} we observed quasinormal modes strongly coupled to a superconducting qubit. These modes had a low Q factor of about 1050, due to large losses for SAW radiation, which determine the dominant channel of energy dissipation for such structures $Q_{i_{PC}}\approx Q_r \ll Q_m.$ 
   
  However, these modes demonstrated a strong coupling with a qubit of $\frac{g}{2\pi}=39$ MHz. Thus, a phononic crystal exhibits significantly higher coupling values than a conventional SAW resonator. This is the result of the acoustic field localization in a small area, which is the main advantage of phononic crystals. Indeed, the length and aperture of a phononic crystal can be much smaller than in a SAW resonator, where large values of ${L_c}$ and $W$ are required to increase the Q factor.
  
    To increase the quality factor of a phononic crystal, we modify its structure. That is shown schematically in Figure 1c. We replace the end parts of the phononic crystal with Bragg mirrors. These mirrors have a period that corresponds to a half wavelength of the fundamental quasinormal mode. Thus, Bragg mirrors reflect the corresponding waves, preventing their radiation from the crystal ends. That suppresses the radiation term expressed by $Q_r$ in (5), and therefore increases $Q_{i_{PC}}$.

\section{Experiment results}\label{sec4}\

  We have carried out an experiment to measure the characteristics of a crystal, which is modified by adding Bragg mirrors to its structure. The measurement circuit is shown in Figure 2(a). A micrograph of the sample is shown in Figure 2(b). The entire phononic crystal consists of  $N = 600$ aluminum strips deposited on a quartz substrate. In the central part of the crystal, there is an IDT consisting of  $N_{IDT} = 100$ strips spaced by a period of $p = 0.95\,\mu$m for stripes with the same polarity. The end parts of that crystal are Bragg mirrors for the main quasinormal modes. Each mirror consists of $N_g = 250$ strips, and their doubled period is $p' = 0.96\,\mu$m. The sample is cooled down to 15 mK in a dilution refrigerator. 
  
\begin{figure}[h]%
\centering
\includegraphics[width=0.9\textwidth]{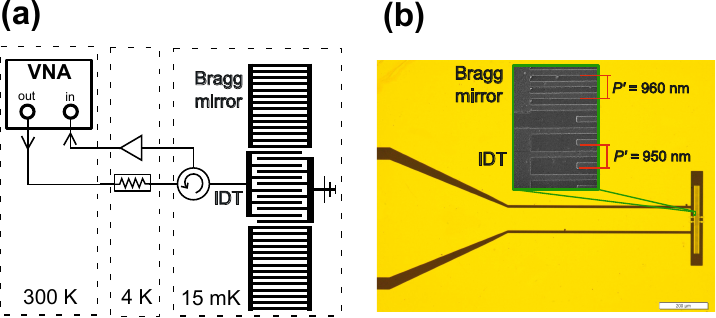}
\caption{\textbf{Measurement circuit and the sample}
\textbf{a. Measurement circuit. } A microwave signal from a VNA goes to the crystal's interdigital transducer (IDT) through an attenuator. The reflected signal goes through the circulator and amplifier back to the VNA.
\textbf{b. Micrograph of the sample.} Optical micrograph of a sample. The inset shows an electron microscope image of the periodic structure details.
}\label{fig2}
\end{figure}

  The microwave signal from a vector network analyzer (VNA) goes to the phononic crystal IDT electrodes through a cascade of attenuators. The reflected radiation returns back through a circulator and a low noise amplifier to the vector network analyzer. The latter measures the complex reflection coefficient $S_{11}(\, f )\,$ of a microwave signal from the phononic crystal at different frequencies $f$.  If the frequency of the applied signal $f$ coincides with the frequency of the quasinormal mode $f_0$, the amplitude of the reflection coefficient undergoes a resonance dip shown in Fig. 3.  
  
\begin{figure}[h]%
\centering
\includegraphics[width=0.6\textwidth]{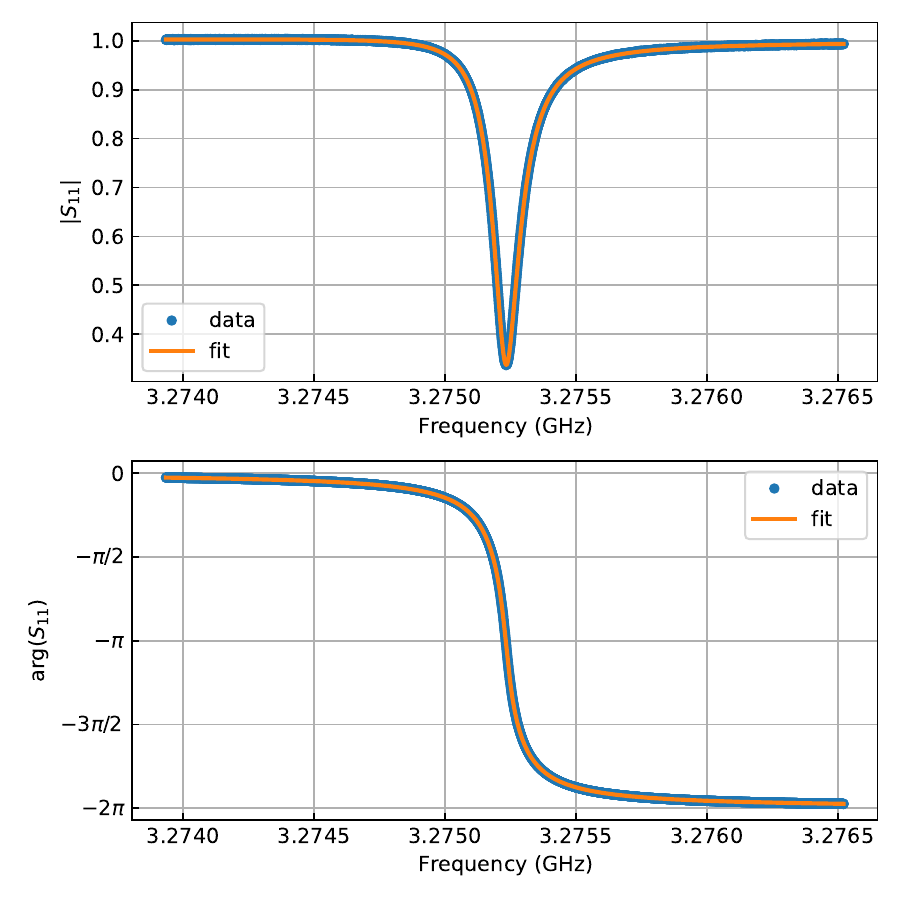}
\caption{\textbf{Microwave signal reflection.} The amplitude (top plot) and phase (bottom plot) of $S_{11}$  near the resonant frequency $f_0$ for a signal applied at $P=-85 dBm$ power.} 
\label{fig3}
\end{figure}

  Near the resonant frequency $f_0$ the reflectance is described by the function \cite{magnusson2015surface}

\begin{equation}
S_{11} =1- \frac{2Q_i/Q_e}{(Q_e+Q_i)/Q_e+i2Q_i (f-f_0)/f_0}e^{i\phi_0},
\label{eq6}
\end{equation}
where $Q_e$ and $Q_i$  are the external and internal  Q-factors, respectively, $\phi_0$ is a parameter accounting for the slight asymmetry of the resonance due to impedance  mismatches.Thus, the experimental values of the Q-factor can be extracted from the reflection coefficient data by it’s fitting according to function (6).

  We have done it for a modified  phononic crystal measured at various temperatures from 100 K to 15 mK. The corresponding plots of $Q_e(T)$ and $Q_{i_{PC}}(T)$ are shown in Figure 4. The internal quality factor grows with decreasing temperature in this temperature range. This behavior can be explained by a decrease in the resistance of aluminum with decreasing temperature, which leads to suppression of losses in the conductor. A qualitatively similar temperature dependence of Q-factor was previously demonstrated for SAW resonators in \cite{magnusson2015surface}. Finally, after decreasing temperature below critical temperature of aluminum ($T_c = 1.18$ K), this dissipation channel disappears and the internal Q-factor reaches its maximum value of about 61000. This value is 60 times higher than the quality factor that we demonstrated earlier \cite{bolgar2020phononic} for a simple phononic crystal without Bragg mirrors.

\begin{figure}[h]%
\centering
\includegraphics[width=0.6\textwidth]{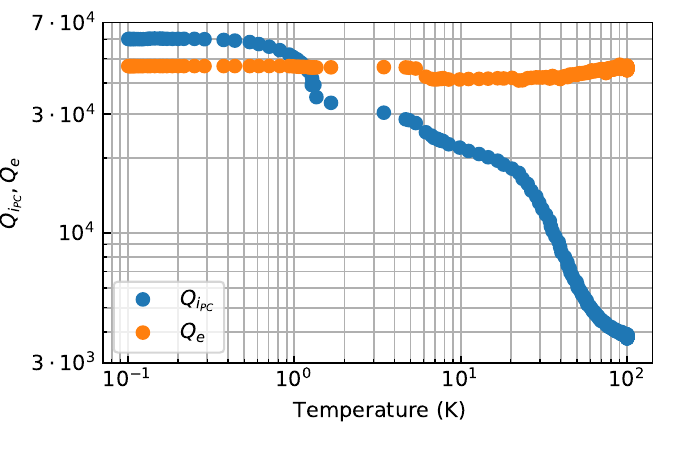}
\caption{\textbf{Temperature dependence of the quality factors of a phononic crystal quasinormal mode.}The internal Q-factor (blue curve) increases with decreasing temperature and reaches a maximum of ${Q_i} \approx 61000$  after the superconducting transition of aluminum strips at  $T < T_c = 1.18\,K$. The external $Q_e$-factor (orange curve) is stable over this temperature range.}
\label{fig4}
\end{figure}

  We also have measured how the internal $Q_{i_{PC}}$-factor depends on the power of the applied signal. The corresponding plot is shown in Figure 5. It demonstrates a decreasing at low powers. This behavior is typical for the single phonon regime and can be explained by the saturation of parasitic two-level systems coupled to a phononic crystal \cite{manenti2016surface}.

\begin{figure}[h]%
\centering
\includegraphics[width=0.6\textwidth]{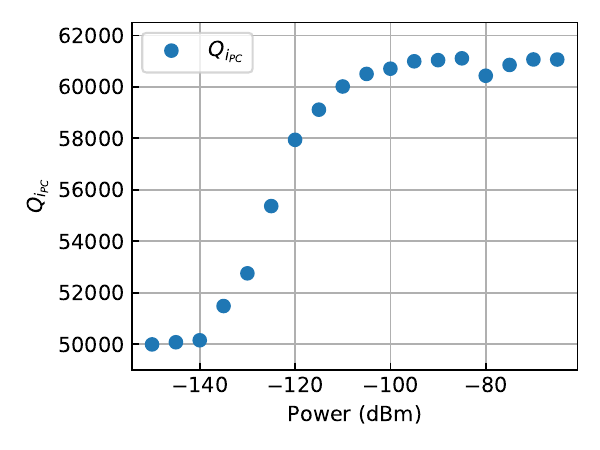}
\caption{\textbf{The dependence of the Q-factor of the phononic crystal quasinormal mode on power.}}
\label{fig5}
\end{figure}

\section{Acoustic field modelling}\label{sec5}\

  For a theoretical description of the observed quality factor increase, we should also simulate the distribution of the acoustic field in the structures of phononic crystals with Bragg mirrors and without them.   We use the reflecting array method (RAM) to simulate a quasinormal modes of the crystal . The equation for one-dimensional surface waves has the form

\begin{equation}
\left[ {\frac{{{\partial ^2}}}{{\partial {x^2}}} - \frac{1}{{v{{(x)}^2}}}\frac{{{\partial ^2}}}{{\partial {t^2}}}} \right]\phi (x,t) = 0,
\label{eq7}
\end{equation}
where $v(x)$ is the velocity of surface waves depending on the  $x$ coordinate. In the region to the left of the phonon crystal $(x < 0)$ there is only a wave going from the crystal to the left and there is no wave coming from the left to the crystal (see Fig. 6.) Similarly, in the region to the right of the crystal ($x > d$)  there is only a wave going to the right. Therefore, the boundary conditions at the ends of this open system can be written as \cite{settimi2003quasinormal, severini2004second}

\begin{equation}
\begin{array}{l}
{v_0}{\partial _x}\phi (x,t) = {\partial _t}\phi (x,t),\,\,\, x < 0, \\
{v_0}{\partial _x}\phi (x,t) =  - {\partial _t}\phi (x,t), \,\,\, x > d,
\end{array}
\label{eq8}
\end{equation}
where $d$ is a structure size,  $v_0$ is a SAW propagation velocity on the free surface. The solution to equation (7)  can be found in the form \cite{severini2004second}

\begin{equation}
\phi (x,t) = \sum\limits_i {{a_i}} {A_i}(x){e^{ - i{\omega _i}t}},
\label{eq9}
\end{equation}
where ${a_i}$ are coefficients of  the expansion in quasinormal modes (QNM),  ${A_i}(x)$ is the spatial distribution of the QNM amplitude, and ${\omega _i}$ is the corresponding complex frequency. This frequency ${\omega _i}$ has a negative imaginary part, which provides the decay of the waves. Substitution of  (9) into (7) leads to the equation for  ${A_i}(x)$     and  ${\omega _i}$ \cite{leung1994completeness,ching1998quasinormal}

\begin{equation}
\left[ {\frac{{{\partial ^2}}}{{\partial {x^2}}} - \frac{{\omega _i^2}}{{v{{(x)}^2}}}} \right]{A_i}(x) = 0.
\label{eq10}
\end{equation}

  The solutions here must also satisfy the condition of mode orthogonality and normalization, which is given by the expression \cite{settimi2003quasinormal}
  
\begin{equation}
\langle A_i\mid A_i \rangle = 2{\omega _i}\int\limits_0^d {\frac{1}{{{v^2}(x)}}A_i^2(x)dx}  + i\frac{1}{{{v_0}}}\left( {A_i^2(0) + A_i^2(d)} \right).
\label{eq11}
\end{equation}
    
    The solution of (10) for each QNM can be represented as the sum of two traveling waves in opposite directions  $A_i(x) = b_i(x) + c_i(x)$.
    
\begin{figure}[h]%
\centering
\includegraphics[width=0.6\textwidth]{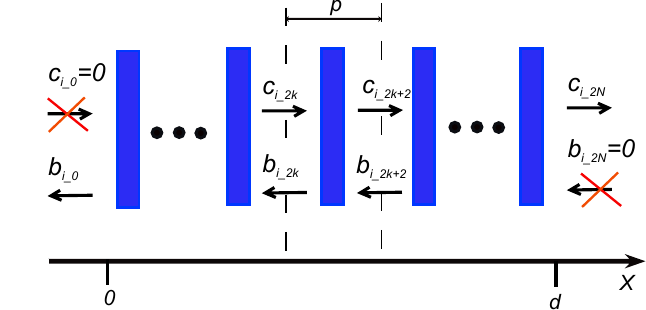}
\caption{\textbf{The scheme of an acoustic field modeling by the reflecting array method.}
The acoustic field is represented by a discrete set of amplitudes at the centers of the stripes (blue rectangles) and between them (dashed lines) for plane waves propagating to the right  $c_i$  and to the left  $b_i$. For a phonon crystal, the boundary conditions assume the absence of waves incident from the left  $c_{i\_0}=0$ and from the right $b_{i\_2N}=0$ . }
\label{fig6}
\end{figure}

We denote the values in the middle of the gaps as ${b_{i\_2k}}$ and  ${c_{i\_2k}}$ and in the middle of the electrodes as ${b_{i\_2k + 1}}$ and ${c_{i\_2k + 1}}$ Then we can write the following recurrence relation for converting amplitudes from even to odd nodes

\begin{equation}
\left( \begin{array}{l}
{b_{i\_2k + 1}}\\
{c_{i\_2k + 1}}
\end{array} \right) = {M_{i\_2k,2k + 1}}\left( 
\begin{array}{l}
{b_{i\_2k}}\\
{c_{i\_2k}}
\end{array} \right).
\label{eq12}
\end{equation}

Here, matrices ${M_{i\_2k,2k + 1}},{M_{i\_2k,2k + 2}}$ have a form \cite{settimi2010classical}

\begin{equation}
{M_{i\_2k,2k + 1}} = \left( {\begin{array}{*{20}{c}}
{\frac{{{E_{i\_h}}{E_{i\_l}}\left( {{n_h} + {n_l}} \right)}}{{2{n_h}}}}&{\frac{{{E_{i\_h}}\left( {{n_h} - {n_l}} \right)}}{{2{n_h}{E_{i\_l}}}}}\\
{\frac{{{E_{i\_l}}\left( {{n_h} - {n_l}} \right)}}{{2{n_h}{E_{i\_h}}}}}&{\frac{{{n_h} + {n_l}}}{{2{n_h}{E_{i\_h}}{E_{i\_l}}}}}
\end{array}} \right),
\label{eq13}
\end{equation}
where  ${E_{i\_l}} = \exp \left( {\frac{{i\,l\,{n_l}{\omega _i}}}{{2{v_0}}}} \right), \,\, {E_{i\_h}} = \exp \left( {\frac{{i\,h\,{n_h}{\omega _i}}}{{2{v_0}}}} \right)$ are the phase factors for the regions where the refractive index ${n_l}$ is low and high ${n_h}$ , respectively. At the boundary of these regions, the reflection and transmission coefficients of waves are determined by expressions $r = \frac{n_h - n_l}{n_h + n_l}, \,\, t = \sqrt {\frac{4n_h n_l }{(n_h + n_l)^2}}, r_s=2r .$ To recalculate the amplitudes from odd to even nodes it is necessary to swap ${n_l}$ and ${n_h}$ in matrix (13).

\begin{figure}[h]%
\centering
\includegraphics[width=0.65\textwidth]{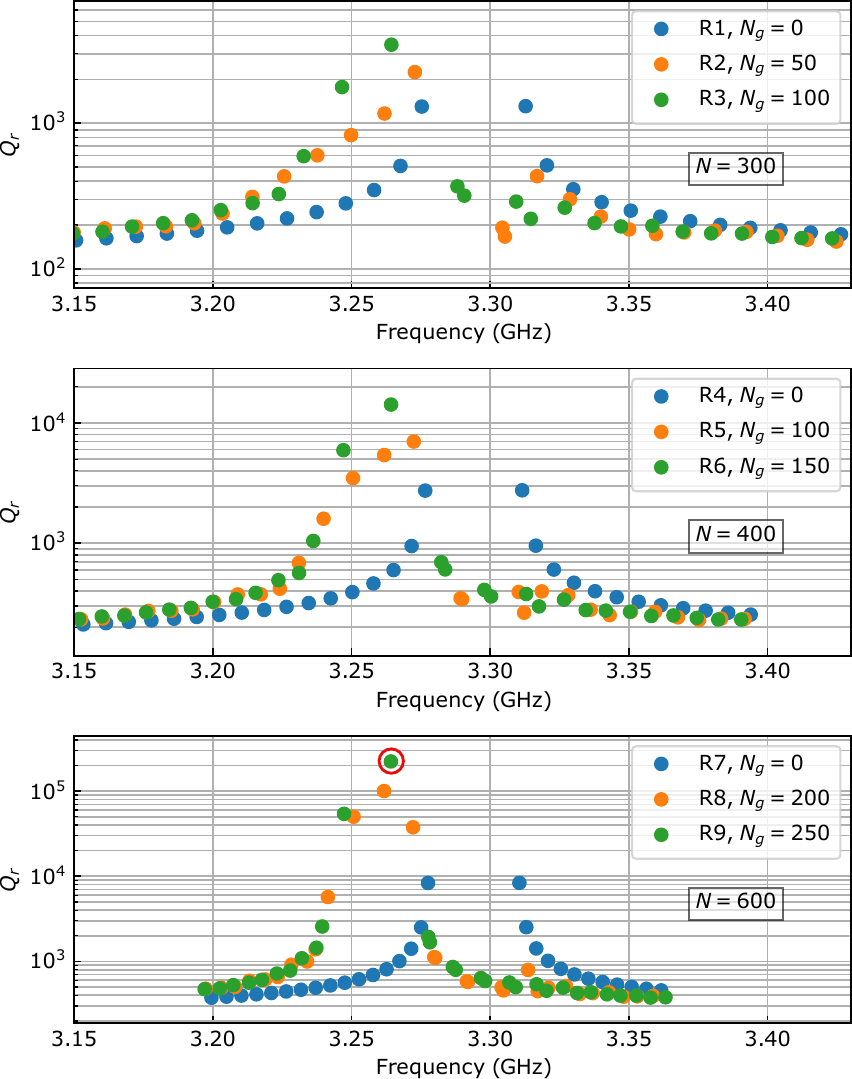}
\caption{\textbf{Calculated Q-factors and frequencies of quasinormal modes of various phononic crystals
a. For short crystals} $N = 300$, ${N_g} = 0$ (blue),  ${N_g} = 50$  (orange), ${N_g} = 100$  (green). 
\textbf{b. For medium crystals} $N = 400$, ${N_g} = 0$ (blue),  ${N_g} = 100$  (orange), ${N_g} = 150$  (green).
\textbf{c. For long crystals} $N = 600$, ${N_g} = 0$ (blue),  ${N_g} = 200$  (orange), ${N_g} = 250$  (green). The point in a red ring corresponds to experimentaly measured mode.  
 }
\label{fig7}
\end{figure}\
  
     It follows from boundary conditions (7) that

\begin{equation}
\begin{array}{l}
c_{i\_0}(\omega _i) = 0,\\
b_{i\_2N}(\omega _i) = 0.
\end{array}
\label{eq14}
\end{equation}

These boundary conditions lead us to the equation for the QNM frequencies

\begin{equation}
\left( {\prod\limits_{k = 0}^{N-1} {{M_{i\_2k + 2,2k}}} } \right)_{2,2} = 0
\label{eq15}
\end{equation}

  By solving numerically equations (15), we obtain a set of frequencies ${\omega _i}$ of quasinormal modes. For each mode, we reconstruct its amplitude at each point using (11) and (12). The modeling method described above considers only one-dimensional propagation of plane waves along the x-axis. It does not take into account the effects of diffraction (diffraction loss) and excitation of bulk modes (bulk loss). Therefore, this method allows us to calculate the theoretical value only for the main contribution to the quality factor ${Q_{{r_i}}}$ from radiation from the edges of the crystal

\begin{equation}
Q_{r_{i}} = \frac{\rm{Re}(\omega _i)}{ - 2\, \rm{Im} (\omega_i)}.
\label{eq16}
\end{equation}
   
\begin{figure}[h]%
\centering
\includegraphics[width=0.6\textwidth]{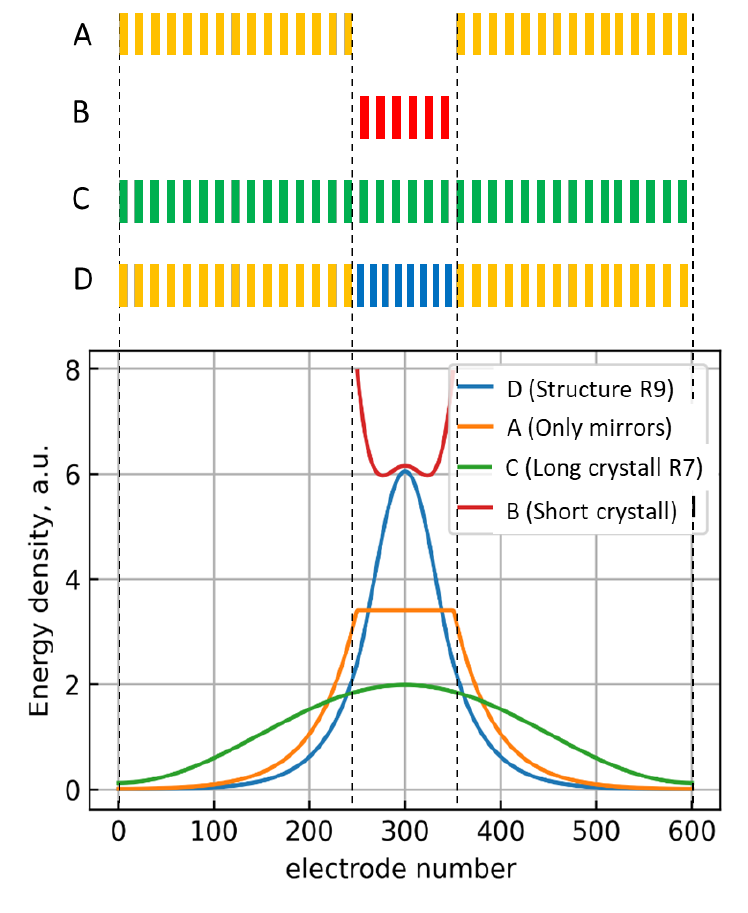}
\caption{\textbf{Distribution of the acoustic field}
The blue curve for R9 structure with Bragg mirrors ($N = 600$, ${N_g} = 250$)  shows a much lower value of the acoustic field at the edges of the crystal and its stronger localization at the centre of the structure than for a homogeneous crystal with  $N = 600$, ${N_g} = 0$ (green curve). Red curve for a short homogeneous crystal, which has a high energy leakage from the ends. Orange curve for an empty cavity between Bragg mirrors demonstrate less field concentration then R9. 
 }
\label{fig8}
\end{figure}

  Figure 7 shows the frequencies and Q-factors of quasinormal modes for nine phononic crystals differing in their parameters $N$ and $N_g$.  We see that the presence of Bragg mirrors at the ends results in a significant increase in the quality factor for phononic crystals of all three considered sizes ($N = 300,\,N = 400,\,N = 600$). In particular, we get  ${Q_{{r_i}}} = 225\,000$ for the structure R9 with  $N = 600$   and  $N_g = 250$, which corresponds to our experimental sample. Taking ${Q_m} \approx 100\,000$ from the work \cite{manenti2016surface} and using (5) we get the theoretical value of the quality factor of 70000, which is in good agreement with the experimentally measured value of 61000.

Figure 8 shows the calculated distribution of the acoustic field amplitude for the fundamental mode in structures with ${N_g} = 600$ and different values of  ${N_g}$ equal to 0 (green curve) and 250 (blue curve). In the first case ${N_g} = 0$ , which means a simple phononic crystal with uniform periodicity. In this case, the field amplitude at the ends of the structure at $x = 0\,$  and  $x = d$ corresponds to significant energy leaks due to the phonon emission. In the second case  ${N_g} = 250$ , and thus 500 of the 600 strips of the structure form Bragg mirrors for the fundamental mode. In this case, the amplitude at the ends of the structure is significantly suppressed, which provides an increase in the quality factor. Our calculations also show the advantage of the modified phononic crystal structure of the R9 type in comparison with an empty SAW cavity (orange curve) and a single crystal without Bragg mirrors (red curve). In the first case, the field concentration in the cavity is lower than in R9, which should lead to a lower coupling value. In the second case, the crystal has too much field at the ends, which sharply reduces its quality factor.

\section{Conclusions}\label{sec6}\

  We have analyzed the main energy dissipation channels in phonon crystals and found that their Q-factor is mainly limited by the emission of phonons from the crystal ends. To suppress this energy leakage, we proposed a modified structure of a phononic crystal with a nonuniform period. The end parts of this structure act as Bragg mirrors for the fundamental mode of the phononic crystal. These mirrors reflect the corresponding SAWs back into the crystal, preventing them from emitting from the ends. Our modeling shows that the internal Q-factor of the fundamental quasinormal mode in the proposed modified phonon crystals can reach ${10^5}$

   We have also fabricated an experimental sample of a phononic crystal with the proposed structure. It has been measured and demonstrates a Q-factor of 61000 at 15 mK. This result is in good agreement with our theoretical predictions. Phononic crystals of this type combine the properties of a high Q-factor and a high concentration of the acoustic field on the chip surface. Due to these properties, modified phononic crystal become a promising element for hybrid devices of quantum acoustodynamics, where a high density of acoustic energy is required to achieve a strong coupling with artificial atoms.

\bmhead{Acknowledgements }\
We acknowledge Russian Science Foundation (Grant No. 21-42-00025) for supporting
the work. This work was performed using technological equipment of MIPT Shared
Facilities Center. 

\bmhead{Competing interests}\
The authors declare no competing interests.

\bmhead{Data availability}\
Relevant data are available from the corresponding author upon request.

\bibliography{Bibliography}% common bib file
\bibliographystyle{ieeetr}


\end{document}